\documentclass[a4paper]{article}

\usepackage{INTERSPEECH2022,multirow,bbding,cite,verbatimbox}

\title{Token-level Speaker Change Detection Using Speaker Difference and Speech Content via Continuous Integrate-and-fire}
\name{Zhiyun Fan$^{1,2}$\thanks{This work is supported by the National Innovation 2030 Major S\&T Project of China under Grant No.2020AAA0104202 and the Key Research Program of the Chinese Academy of Sciences under Grant No.ZDBS-SSW-JSC006 and Strategic Priority Research Program of the Chinese Academy of Sciences under Grant No.XDA27030300.}, Zhenlin Liang$^3$, Linhao Dong$^3$, Yi Liu$^3$, Shiyu Zhou$^1$, \\ 
      Meng Cai$^3$, Jun Zhang$^3$, Zejun Ma$^3$, Bo Xu$^1$}
\address{
  $^1$Institute of Automation, Chinese Academy of Sciences, China\\
  $^2$School of Artificial Intelligence, University of Chinese Academy of Sciences, China\\
  $^3$Bytedance AI LAB}

\email{fanzhiyun2017@ia.ac.cn, \{liangzhenlin.lzl, donglinhao\}@bytedance.com}

\begin{document}

\maketitle
\begin{abstract}
In multi-talker scenarios such as meetings and conversations, speech processing systems are usually required to segment the audio and then transcribe each segmentation. These two stages are addressed separately by speaker change detection (SCD) and automatic speech recognition (ASR). Most previous SCD systems rely solely on speaker information and ignore the importance of speech content. In this paper, we propose a novel SCD system that considers both cues of speaker difference and speech content. These two cues are converted into token-level representations by the continuous integrate-and-fire (CIF) mechanism and then combined for detecting speaker changes on the token acoustic boundaries. We evaluate the performance of our approach on a public real-recorded meeting dataset, AISHELL-4. The experiment results show that our method outperforms a competitive frame-level baseline system by 2.45\% equal coverage-purity (ECP). In addition, we demonstrate the importance of speech content and speaker difference to the SCD task, and the advantages of conducting SCD on the token acoustic boundaries compared with conducting SCD frame by frame.

\end{abstract}
\vspace{-0.1cm}
\section{Introduction}
\vspace{-0.1cm}
Speaker change detection (SCD) is the task of locating the time points when a different speaker starts to speak. The SCD system is often applied as a submodule of speaker diarization, which determines “who spoke when” or is used as a front-end of automatic speech recognition (ASR). The performance of SCD will greatly affect the subsequent models.

Many different SCD systems have been proposed in the literatures. 
The distance-based methods \cite{chen1998speaker,siegler1997automatic,tran2011comparing,cettolo2005evaluation,gish1991segregation} are studied earliest, which calculate the distance between features in the adjacent windows, and once the distance exceeds the threshold, a speaker change is detected. In addition, the model-based methods \cite{ge2017speaker,anguera2012speaker,malegaonkar2007efficient,castaldo2008stream} segment the input audio into fix-length segments that are assumed to contain only one speaker, and models are trained to extract speaker embeddings for each segment. Then the distance between adjacent speaker embeddings is used to decide whether speaker change happens. More recently, some end-to-end SCD systems \cite{yin2017speaker,sari2019pre,sari2020auxiliary,hruz2017convolutional} are proposed to predict the speaker change at the end of the neural network without relying on a distance metric. Yin et al. address the SCD as a sequence labeling task, which predicts each frame as a change point or not \cite{yin2017speaker}. Leda et al. extract speaker embeddings for adjacent segments with a shared siamese network, then a fully connected (FC) classifier decides whether there is a speaker change between these two segments \cite{sari2019pre}. Marek et al. propose a Convolutional Neural Networks (CNN) based system to the SCD task \cite{hruz2017convolutional}. 

The SCD systems mentioned above only rely on acoustic information such as pitch \cite{hogg2019speaker}, speaker embeddings \cite{sari2019pre}, and silence \cite{kemp2000strategies}. However, the speech content is another critical cue for the SCD task \cite{li2009improving}. There has been little effort in exploiting speech content for the SCD task. 
Meng et al. formulate text-based SCD as a binary sentence-pair classification problem, which judges whether the speaker is changing between consecutive sentence pairs \cite{meng2017hierarchical}. 
Jin et al. investigate the way to integrate content and acoustic cues with sequence to sequence model to improve speaker diarization performance \cite{park2018multimodal}. India et al. prove that the combination of content and acoustic features outperforms the cosine distance-based baseline where only acoustic data is used \cite{india2017lstm}. Anidjar et al. propose a hybrid framework for the SCD problem that is learned by content information and speech signals \cite{anidjar2021hybrid}. Most of the existing methods combining speaker and content information rely on the acoustic boundaries from the manual annotation or the forced-alignment results produced by the ASR, which leads to complex processes and time-consuming annotations.


In this paper, we propose an end-to-end SCD system that combines speech content and speaker difference to detect speaker changes between tokens. We construct a joint model for ASR, speaker identification (SID) and SCD. The ASR part follows the continuous integrate-and-fire (CIF) based encoder-decoder \cite{dong2020cif} structure, which can generate speech representations and token acoustic boundaries. The SID part uses the located token acoustic boundaries by the CIF to generate token-level speaker embeddings. The SCD part calculates the speaker difference between the token-level speaker embeddings and then combines speaker difference with speech content to detect speaker change.
We train and evaluate the proposed model on a public real-recorded dataset, AISHELL-4. The experiment results of our method show a performance improvement compared to a competitive open-sourced frame-level baseline. Further experiments demonstrate the advantages of using speaker difference and speech content for the SCD task.




\begin{figure*}[!ht]
  \centering
  \includegraphics[width=0.8\linewidth]{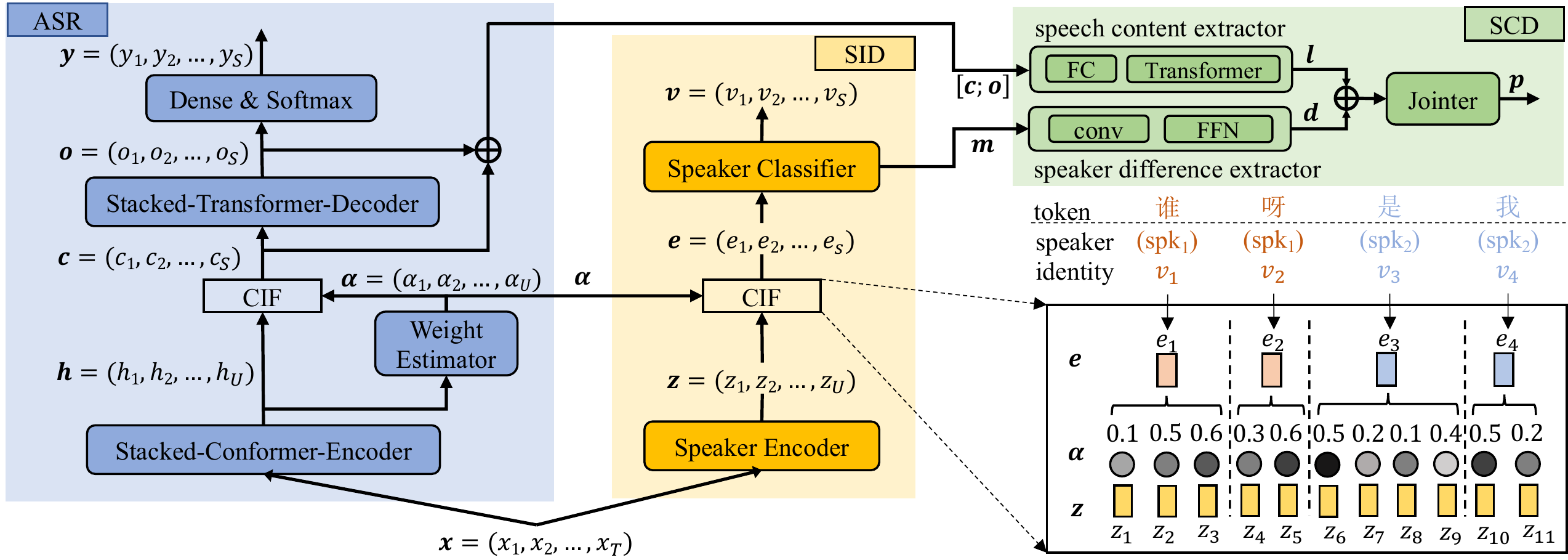}
  \vspace{-0.3cm}
  \caption{Schematic diagram of the proposed model. The ASR and speaker encoder have extra pre-training. During the joint training of the ASR, SID and SCD, the parameters of ASR are frozen. The lower right corner of the figure uses an example to show how the CIF converts the frame-level speaker representations into token-level speaker representations. The dashed line in the box represents the token acoustic boundaries. The token-level speaker representations $e_1=0.1*z_1+0.5*z_2+0.4*z_3$, $e_2=0.2*z_3+0.3*z_4+0.5*z_5$, $e_3=0.1*z_5+0.5*z_6+0.2*z_7+0.1*z_8+0.1*z_9,e_4=0.3*z_9+0.5*z_{10}+0.2*z_{11}$.}
  \label{fig:joint}
  \vspace{-0.6cm}
\end{figure*}
\vspace{-0.1cm}
\section{CIF-based ASR}
\vspace{-0.1cm}
\label{cifasr}
The continuous integrate-and-fire (CIF) based ASR follows an encoder-decoder framework connected by the CIF mechanism, a soft and monotonic alignment \cite{dong2020cif}. The blue part in Fig.\ref{fig:joint} illustrates the structure of the CIF-based ASR. The encoder transforms the acoustic feature sequence $\boldsymbol{x}=(x_1,x_2,...,x_T)$ to a encoded output $\boldsymbol{h}=(h_1,h_2,...,h_U)$, where $T$ and $U$ are the total length of the input sequence and the encoded output, respectively. 
A weight estimator calculates the information weights $\boldsymbol{\alpha}=(\alpha_1,\alpha_2,...,\alpha_U)$ for $\boldsymbol{h}$ at each time step. $\alpha_u$ scales the amount of information contained in $h_u$. In the training stage, a scaling operation is applied to the raw weights $\boldsymbol{\alpha}$ to ensure that the sum of the weights $\boldsymbol{\alpha}$ is equal to the length of targets $\boldsymbol{\Bar{y}}=(\Bar{y}_1,\Bar{y}_2,...,\Bar{y}_S)$.
\begin{equation}
\setlength{\abovedisplayskip}{3pt} 
\setlength{\belowdisplayskip}{3pt}
\label{eq0}
   \boldsymbol \alpha' = \frac{S}{\sum_{u=1}^U\alpha_u} \cdot \boldsymbol \alpha 
\end{equation}

The CIF processes the frame-level speech representations $\boldsymbol{h}$ and information weights $\boldsymbol \alpha'$ in a left-to-right manner. The information weight $\alpha_u'$ is accumulated to $\alpha_u^a$. Meanwhile, the token-level speech representations are integrated frame by frame.
\begin{equation}
\setlength{\abovedisplayskip}{3pt} 
\setlength{\belowdisplayskip}{3pt}
\label{eq1}
\alpha_{u}^a=\alpha_{u-1}^a+\alpha_u'
\end{equation}
\begin{equation}
\setlength{\abovedisplayskip}{3pt} 
\setlength{\belowdisplayskip}{3pt}
\label{eq2}
h_u^{a} = h_{u-1}^a + \alpha_u' \cdot h_u
\end{equation}


Once the accumulated weight $\alpha_{u}^a$ reaches a threshold $\beta$, a token-level acoustic boundary is located, and these time steps are called fired time steps, denoted as $u^{*}$. At each fired time step, the information weight $\alpha_{u^*}$ is divided into two parts, $\alpha_{u1}$ and $\alpha_{u2}$. $\alpha_{u1}$ is used for completing the current integration of token-level representation $c_i$. The sum of information weights consumed in a complete integration is $1$. $\alpha_{u2}$ remains for the next integration. A complete integration spans from the last fired time step to the current fired time step.
\begin{equation}
\setlength{\abovedisplayskip}{3pt} 
\setlength{\belowdisplayskip}{3pt}
\label{eq4}
    \alpha_{u1} = 1 - \alpha_{u^*-1}^a
\end{equation}
\begin{equation}
\setlength{\abovedisplayskip}{3pt} 
\setlength{\belowdisplayskip}{3pt}
\label{eq5}
    \alpha_{u2} = \alpha_{u^*}' - \alpha_{u1}
\end{equation}
\begin{equation}
\setlength{\abovedisplayskip}{3pt} 
\setlength{\belowdisplayskip}{3pt}
\label{eq6}
    c_{i} = h_{u^*-1}^a + \alpha_{u1}\cdot h_{u^*}
\end{equation}

After firing the current integration results $c_i$ (Eq.\,\ref{eq6}), the accumulated weight and token-level representation state will be reset as follows:
\begin{equation}
\setlength{\abovedisplayskip}{3pt} 
\setlength{\belowdisplayskip}{3pt}
\alpha_{u^*}^a=\alpha_{u2}
\end{equation}
\begin{equation}
\setlength{\abovedisplayskip}{3pt} 
\setlength{\belowdisplayskip}{3pt}
\label{eq8}
h_{u^*}^{a} = \alpha_{u2} \cdot h_{u^*}
\end{equation}

The process described from Eq.\,\ref{eq1} to Eq.\,\ref{eq8} will be repeated until the last time step $U$. Then the CIF finishes transferring the encoded output $\boldsymbol h$ into token-level representations $\boldsymbol{c}=(c_1,c_2,...,c_S)$, and these fired time steps are located as token acoustic boundaries. Since we apply the scaling operation in Eq.\,\ref{eq0} and the threshold $\beta$ is set to $1$, the length of $\boldsymbol{c}$ is equal to the length of target labels $\boldsymbol{\Bar{y}}$. For the convenience of the following description, we summarize the process of integration and fire formulated from Eq.\,\ref{eq0} to Eq.\,\ref{eq8} as follow:
\begin{equation}
\setlength{\abovedisplayskip}{3pt} 
\setlength{\belowdisplayskip}{3pt}
    \boldsymbol{c} = \text{CIF}(\boldsymbol{h}, \boldsymbol \alpha)
\end{equation}

The decoder receives the fired token-level representations $\boldsymbol{c}$ and predicts the label sequence $\boldsymbol{\Bar{y}}=(\Bar{y}_1,\Bar{y}_2,...,\Bar{y}_S)$ in an autoregressive manner. 
\begin{equation}
\setlength{\abovedisplayskip}{3pt} 
\setlength{\belowdisplayskip}{3pt}
    \text{Pr}(y_i | \Bar{y}_{1:i-1},c_{1:i}) = \text{ASRDecoder}(\Bar{y}_{i-1},c_i)
\end{equation}

During inference, the scaling operation in Eq.\,\ref{eq0} is skipped. To alleviate this mismatch between training and inference, the quantity loss and tail handling are adapted. For more details of the CIF-based ASR, we refer readers to \cite{dong2020cif}.


\vspace{-0.1cm}
\section{Proposed method}
\label{method}
\vspace{-0.1cm}
\subsection{Overview}
\vspace{-0.1cm}
In this paper, we propose to use speaker difference and speech content to conduct a token-level speaker change detection (SCD). The model body is a joint framework for automatic speech recognition (ASR), speaker identification (SID) and SCD. The three parts are distinguished by different colors in Fig. \ref{fig:joint}. The ASR part transfers the input feature sequence $\boldsymbol{x}$ to the target token sequence $\boldsymbol{y}$ while providing the token acoustic boundaries contained in the information weights $\boldsymbol \alpha$ and speech representations $\boldsymbol{c}$ and $\boldsymbol{o}$. The SID part conducts a token-level speaker classification, and then generates token-level speaker embeddings $\boldsymbol{m}$. The SCD part captures the speaker difference $\boldsymbol{d}$ between the speaker embeddings $\boldsymbol{m}$, and combines it with the speech content $\boldsymbol{l}$ to detect speaker changes between tokens.
\vspace{-0.1cm}
\subsection{Model Structure}
\vspace{-0.1cm}
The details of the model body is shown in Fig.\ref{fig:joint}. Given the input feature sequence $\boldsymbol{x}=(x_1,x_2,...,x_T)$, the CIF-based ASR predicts the token sequence as described in Section \ref{cifasr}. 

The SID consists of a speaker encoder, a classifier, and the CIF module. The speaker encoder transfers $\boldsymbol{x}$ to frame-level speaker representations $\boldsymbol{z}=(z_1,z_2,...,z_U)$, and it has the same temporal downsampling as the ASR encoder to make the length of their output sequence consistent.
Then the CIF mechanism uses the frame-level speaker representations $\boldsymbol{z}$ and the information weights $\boldsymbol \alpha$ provided by the ASR part to generate token-level speaker representations $\boldsymbol{e}$.
\begin{equation}
\setlength{\abovedisplayskip}{3pt} 
\setlength{\belowdisplayskip}{3pt}
    \boldsymbol{e} = (e_1,e_2,...,e_S)=\text{CIF}(\boldsymbol{z},\boldsymbol \alpha)
\end{equation}

In the calculation process of the CIF, the fired time steps and the length of output sequence $S$ are dependent on the information weights $\boldsymbol \alpha$, so the token-level speaker representations and speech representations can be one-by-one correspondence. 
The speaker classifier stacks two full-connected (FC) layers. The first layer transfers $\boldsymbol{e}$ to the token-level speaker embeddings $\boldsymbol{m}=(m_1,m_2,...,m_S)$. The second layer projects $\boldsymbol{m}$ to the speaker probability distribution $\boldsymbol{v}=(v_1,v_2,...,v_S)$.

The SCD part consists of a speech content extractor (SCE), a speaker difference extractor (SDE), and a jointer. The SCE stacks a FC layer and uni-directional transformer layers with masked multi-head self-attention \cite{nicolson2020masked}. It pre-processes the speech presentations from the CIF output $\boldsymbol{c}$ and ASR decoder output $\boldsymbol{o}$, and generates the speech content $\boldsymbol{l}=(l_1,l_2,...,l_S)$. The SDE stacks a 1d-convolution layer and a feed-forward network (FFN) layer to calculate the speaker difference $\boldsymbol{d}=(d_1,d_2,...,d_S)$ between the token-level speaker embeddings $\boldsymbol{m}$. Using 1D convolution with kernel size 3, the SDE calculates the difference between the speaker embeddings of the last token, the current token and the next token. Then the speaker difference and the speech content are concatenated and fed into the jointer. The FC layer based jointer combines these two cues to predict speaker change probability $\boldsymbol{p}=(p_1,p_2,...,p_S)$.

\vspace{-0.2cm}
\subsection{Training}
\vspace{-0.1cm}
Before the joint training of the proposed model, we pre-train the CIF-based ASR part and speaker encoder, respectively. The loss used in the pre-training of ASR is consistent with our previous work \cite{dong2020cif}.
The pre-training of the speaker encoder is an utterance-level speaker classification. We add an extra average pooling and an output FC layer upon the speaker encoder to perform the pre-training. The AMSoftmax \cite{liu2019large} loss is used to capture more discriminative speaker embeddings.

During the joint training stage, the parameters of ASR part are frozen. The loss is the interpolation of a token-level AMSoftmax loss for speaker classification and a token-level binary cross-entropy (BCE) loss for speaker changes detection.
\begin{equation}
\setlength{\abovedisplayskip}{3pt} 
\setlength{\belowdisplayskip}{3pt}
\mathcal{L}_{\text{joint}}= \text{AMS}(\boldsymbol{v},\boldsymbol{\Bar{v}}) + \text{BCE}(\boldsymbol{p},\boldsymbol{\Bar{p}})
\end{equation}
where $\boldsymbol{\Bar{v}}$ and $\boldsymbol{\Bar{p}}$ are token-level speaker identity label and speaker change label, respectively.

  

\vspace{-0.1cm}
\section{Experiments}
\label{experiments}
\vspace{-0.1cm}
\subsection{Dataset}
\vspace{-0.1cm}
AISHELL-4 \cite{fu2021aishell} is a public real-recorded conversation speech dataset in conference scenario. It contains $118$ hours of audio recorded by an 8-channel microphone array. All our experiments only use the first channel. Each of the total $211$ meeting sessions contains $4$ to $8$ speakers. As the speech recognition for the overlapped speech is beyond the scope of this paper, we exclude the overlapped speech of the AISHELL-4. The excluded rules are that (1) one to four consecutive intervals annotated in the original annotation file form a sentence. (2) these sentences containing overlapped speech for more than $1$s will be discarded. (3) these sentences containing overlapped speech that accounts for more than 10\% of the duration of any interval will be discarded. (4) these sentences containing more than $10$s silence will be discarded. The training and test set are processed independently. Then we extract five meeting sessions as the development set from the original training set, and the remainder is used as our training set. The test set is further excluded these sentences containing no speaker change. The details of the processed data are shown in Table \ref{table:data}. In order to facilitate readers to reproduce our work, we release the data processing scripts\footnote{https://github.com/zhiyunfan/aishell4-preprocess}.
\begin{table}[t]
  \caption{Details of the processed data.}
  \vspace{-0.2cm}
  \centering
  \addvbuffer[-2pt -2pt]{
  \begin{tabular}{lccc}
    \toprule
  & \textbf{Train} & \textbf{Dev} &\textbf{Test}  \\
    \midrule
    \#Samples & 84705 & 3586 & 1971 \\
    \#Speakers & 36 & 19 & 25 \\
  Min duration (s) & 0.11 & 0.19 & 0.61 \\
   Mean duration (s) & 7.90  & 7.68 & 11.75  \\
   Max duration (s) & 71.87 & 46.28  &  60.78 \\
    Total duration (h)& 185.79 & 7.65 &  6.43 \\
    \bottomrule
  \end{tabular}}
  \label{table:data}
\end{table}

\vspace{-0.2cm}
\subsection{Evaluating Metric}
\vspace{-0.1cm}
All our experiments are evaluated on the equal coverage-purity (ECP). The formula of the purity and the coverage \cite{bredin2017pyannote} are as follows:
\begin{equation}
\setlength{\abovedisplayskip}{3pt} 
\setlength{\belowdisplayskip}{3pt}
{\rm Purity}=\frac{\sum_{h \in H}\text{max}_{r \in R}|h \cap r|}{\sum_{h \in H}|h|}
\label{purity}
\end{equation}
\begin{equation}
\setlength{\abovedisplayskip}{3pt} 
\setlength{\belowdisplayskip}{3pt}
{\rm Coverage}=\frac{\sum_{r \in R}\text{max}_{h \in H}|r \cap h|}{\sum_{r \in R}|r|}
\label{coverage}
\end{equation}
where $R$ and $H$ are the set of segments cut by the reference and hypothesized change points, respectively. These points exceed a threshold $\theta$ are detected as speaker changes. $|s|$ is the duration of segment $s$. $r \cap h$ is the intersection of segments $r$ and $h$. 
\vspace{-0.1cm}
\subsection{Experimental Setup}
\vspace{-0.1cm}
We extract input features using the same setup as \cite{dong2020cif}. A convolutional layer with $1/2$ temporal downsampling is used as the front-end of the ASR, and the filter number is set to $64$. The ASR encoder consists of $15$ Conformer \cite{gulati2020conformer} encoder layers with 8 attention heads, $400$ attention dimensions and $1600$ FFN dimensions. There is a max-pooling layer for $1/2$ temporal downsampling after the fifth and tenth layers, respectively. The weight estimator consists of a 1-dimensional convolutional layer and a FC layer. The kernel size of the convolutional layer is set to $3$, and the filter number is $400$. The FC layer has one output unit with sigmoid activation. The ASR decoder consists of $2$ Transformer \cite{dong2020cif} layers with $8$ attention heads, $400$ attention dimensions and $1600$ FFN dimensions. The speaker encoder has the same architecture as the 18-layer ResNet in \cite{he2016deep} except for not the final average pooling layer and halving the channels of all the convolutional layers. Both the speaker encoder and the ASR encoder have a $1/8$ temporal downsampling. The speaker decoder stacks two FC layers. The first layer has 128 units with ReLU activation, the second layer has $36$ (total speaker number in the training set) output units with softmax nonlinearity. The speech content extractor consists of an FC layer and transformer layers. The FC layer projects the inputs to $400$ dimensions. The structure of transformer layers is consistent with the ASR decoder. The speaker difference extractor consists of a 1d-convolutional layer with kernel size $3$ and an FFN layer with $512$ hidden dimensions. The jointer consists of a hidden FC layer with $528$ units and an output FC layer with $1$ unit. The threshold $\beta$ in these two CIF is set to $1$.

For the pre-training of ASR, 
the learning rate warms up for the first $1$k iterations to a peak of $10^{-3}$ and holds on for the next $40$k iterations, and then linearly decays for the remainder. 
For the pre-training of the speaker encoder, the SGD optimizer is used. The learning rate warms up for the first $1$k iterations to a peak of $10^{-4}$ and holds on for the next $5$k iterations.
During the joint training stage, 
the learning rate warms up for the first $1$k iterations to a peak of $10^{-3}$, and holds on for the remainder. 

\vspace{-0.1cm}
\subsection{Results}
\vspace{-0.1cm}
\begin{table}[th]
\vspace{-0.2cm}
  \caption{ECP/\% for baseline systems and the proposed method.}
  \vspace{-0.2cm}
  \centering
  \begin{tabular}{c|c|c|c|c}
    \toprule
\textbf{Exp} & \multicolumn{3}{c|}{\textbf{Model}}  & {\textbf{ECP}}  \\
    \midrule
 A1 &   \multicolumn{3}{c|}{BSL with Bi-LSTM \cite{bredin2020pyannote}} & 84.05      \\
 A2 &   \multicolumn{3}{c|}{BSL with ResNet18} &  85.77 \\
    \midrule
    & \textbf{Speaker} & \textbf{Difference} & \textbf{Speech content} & \\ 
    \midrule
B1  & $\boldsymbol{z}$ & $\times$ & $\times$  & $82.08$  \\
B2  &  $\boldsymbol{z}$  & \checkmark & $\times$ & $82.18$ \\
B3  &  $\boldsymbol{m}$ & $\times$  & $\times$  & $81.47$ \\
B4  &  $\boldsymbol{m}$ & \checkmark  &  $\times$ &  $86.99$ \\
B5  &  $\times$ & $\times$ & \checkmark & 86.73 \\
B6  &   $\boldsymbol{m}$ & \checkmark & \checkmark &\textbf{88.22}  \\
\midrule
 & \multicolumn{3}{c|}{\textbf{Convolution context}} &  \\
 \midrule
 C1&  \multicolumn{3}{c|}{$[-1,1]$} & 88.22 \\
 C2& \multicolumn{3}{c|}{$[-2,2]$} & 87.61  \\
 C3& \multicolumn{3}{c|}{$[-3,3]$} & 87.99 \\
    \bottomrule
  \end{tabular}
  \label{table1}
\end{table}
\vspace{-0.45cm}
\subsubsection{Baseline results}
\vspace{-0.1cm}
We build two baseline systems with the Pyannote1.1 toolkit \cite{bredin2020pyannote}. The baseline system A1 with Bi-LSTM follows the structure in \cite{yin2017speaker}, which addresses the SCD as a binary sequence labeling (BSL) task. The input MFCC feature sequence is encoded by the Bi-LSTM. Then a multi-layer perceptron (MLP) projects the encoded sequence to a frame-level score sequence between $0$ and $1$. For a fair comparison, the baseline A2 replaces the Bi-LSTM with the ResNet18, the structure used in our speaker encoder. The results of these two baseline systems are shown in the A1 and A2 of Table \ref{table1}.
\vspace{-0.2cm}
\subsubsection{Evaluation with token-level speaker difference}
\vspace{-0.1cm}
The B1 to B4 in Table \ref{table1} are four variants of the proposed model. All the four experiments ablate the use of the speech content. B1 feeds the frame-level speaker representations $\boldsymbol{z}$ into the SDE, and skips the convolutional layer in the SDE. Compared with B1, B2 uses the convolutional layer in the SDE to capture the speaker difference. In contrast, B3 and B4 feed the token-level speaker embeddings $\boldsymbol{m}$ into the SDE, and then the jointer detects speaker changes on the acoustic boundaries. 

Comparing B2 with B1, we find that using convolution in the SDE to capture the speaker difference between the frame-level speaker representations $\boldsymbol{z}$ brings little improvement. Comparing B4 with B3, both of which use the token-level speaker embeddings $\boldsymbol{m}$, the use of speaker difference brings a $5.52$\% ECP improvement. These results indicate that calculating the speaker difference between token-level speaker embeddings is more useful to the SCD task. Moreover, compared with the frame-by-frame speaker change detection, detecting speaker changes on the token acoustic boundaries can greatly reduce the number of candidates to be processed. From the results of B1 to B4, we could find that the token-level speaker difference information (B4) obtains the best performance. Thus, all subsequent experiments use token-level speaker difference for the token-level SCD.

\begin{figure}[ht]
  \centering
  \includegraphics[width=0.7\linewidth]{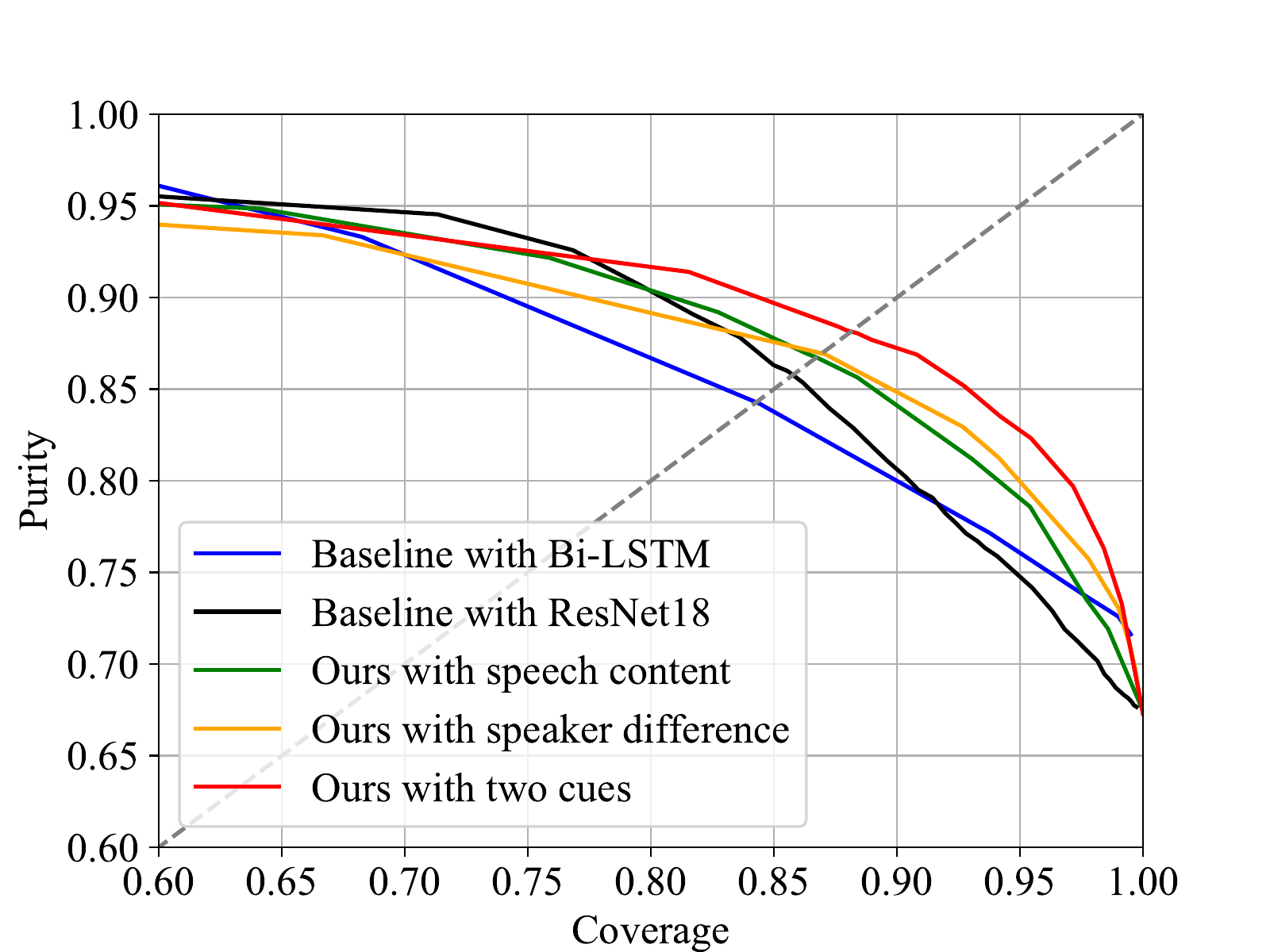}
  \vspace{-0.2cm}
  \caption{Coverage-purity measured on baseline methods and our method with speaker difference and speech content.}
  \vspace{-0.4cm}
  \label{curve}
\end{figure}
\vspace{-0.2cm}
\subsubsection{SCD with the speaker difference and the speech content}
\vspace{-0.1cm}
Another advantage of using token-level speaker difference is the ability to combine it with token-level speech content at the same granularity. B6 in Table \ref{table1} shows the result of the proposed method with both token-level speaker difference and speech content. The results of B4 and B5 represent that our method ablates speech content or speaker difference, respectively. We find that using one of the speaker difference or speech content in our method has outperformed these two baseline systems (A1 and A2), and B6 shows that combining these two cues can bring further performance improvement, and outperforms the baseline system with ResNet18 by $2.45$\% ECP.

In Fig \ref{curve}, we plot the coverage-purity curves of B4, B5, B6 and these two baseline systems (A1 and A2). All curves are obtained by varying the threshold $\theta$. The dashed line is diagonal, where purity is equal to coverage. Among the five systems, area under the curve of B6 is the largest, and B6 performs better in most of the varing thresholds.

\vspace{-0.2cm}
\subsubsection{Evaluation with different speaker context settings}
\vspace{-0.1cm}
In the previous experiments, all convolutional layers in the SDE have the context of $[-1,1]$, which means that the model refers to the speaker embedding of the previous token and the next token when determining whether there is a speaker change between current token and next token. We then attempt to improve the proposed method by more speaker contexts. The results of our method with different speaker contexts are shown from C1 to C3 in Table \ref{table1}. The best ECP performance is achieved by the $[-1,1]$ context. After our analysis of the corpus, we find a lot of speaker change points where one person says only one token inside another person's speech. For such cases, $[-1,1]$ context is enough for the SCD, and more contexts bring confusion. 

\vspace{-0.1cm}
\section{Conclusions}
\label{conclusion}
\vspace{-0.1cm}
In this paper, we propose to use both speaker difference and speech content to detect speaker changes on the located token acoustic boundaries by the CIF.
The experiments demonstrate the advantages of using the speaker difference and the speech content for the SCD task, and show performance improvement of detecting speaker changes on the token acoustic boundaries. Specifically, our token-level SCD system using both speaker difference and speech content outperforms the competitive frame-level baseline by $2.45$\% ECP.
In the future, using the CIF to integrate the other information contained in speech with content for more speech processing tasks will be a potential research direction.
\vspace{-0.1cm}
\section{Acknowledgments}
\vspace{-0.1cm}
The authors wish to thank Yangcheng Wu for his assistance with the model training when he was an intern in Bytedance AI Lab.

\bibliographystyle{IEEEtran}

\bibliography{mybib}


\end{document}